# GENDER IN VIDEO GAMES: CREATIVITY FOR INCLUSIVITY

*María Isabel RIVAS GINEL*
Université de Bourgogne Franche-Comté & Universidad de Valladolid
ORCID : 0000-0001-9509-684X

*Sarah THEROINE*
Université de Bourgogne Franche-Comté
ORCID : 0000-0003-4238-763X

Video game localisation, a field highly impacted by the lack of visual environment and text linearity, forces translators to create inclusive solutions in terms of gender to overcome the hurdles created by variables. This paper will introduce the specificities of this sector and present an analysis of some of those techniques extracted from parallel corpora compiled from video games that include female, transgender, non-binary, and non-sexualised characters.

## LE GENRE DANS LES JEUX VIDÉO : LA CRÉATIVITÉ POUR L'INCLUSIVITÉ

La localisation de jeux vidéo, un domaine très impacté par l'absence d'accès au jeu ainsi que par le manque de linéarité textuelle, oblige les traducteurs à trouver des solutions inclusives pour faire face aux problèmes créés par les variables. Cet article se propose d'introduire les caractéristiques du secteur ainsi que de présenter l'analyse des techniques extraites des corpus parallèles compilés à partir des jeux ayant des personnages féminins, transgenres, non-binaires, et non-sexualisés.
**Mots-clés :** *genre, jeux vidéo, corpus, linguistique appliquée, transcréation, localisation, inclusivité.*

## Video game localisation: origins and transcreation

Video game localisation remains remarkably young compared with other translation domains and, as part of the multimedia translation branch, its birth and development are intrinsically linked to the appearance of computers, the implementation of GILT practices (globalisation, internationalisation, localisation, and translation) and technological advances in general. Similarly to utility software, video game localisation evolved following a trial and error process that led to today's practices. For instance, the democratisation of personal computers and the emergence of an international market soon led to the realisation that "it was essential to convert the software so that users saw a product in their own language and firmly based in their own culture" [1, p. x]. These cultural differences transcended linguistic matters and included aspects related to formats, dates, "character sets for the digital representation of writing systems, encodings to enable the storage and retrieval of data in languages other than English, collation rules, [...] as well as calendars and decimal separators (period or comma)" [2, p. 148]. Moreover, the development and implementation of the GILT process in the field of localisation — and especially internationalisation practices — allowed companies to prepare the product for localisation in advance, reduced the need for *ad-hoc* modifications to the programme, and eliminated the necessity of dealing with separate versions of the same product depending on the language. Even though these improvements reduced the number of technical difficulties that affected translators, their workflow remained largely techno-centric and, as noted by Pym [3], the environment in which they work was created around CAT tools.

Thus, the professionals are deprived of a substantial part of the context and forced to deal with segments (referred to as 'strings' in the industry) instead of paragraphs or even whole documents.

Although utility software, mobile apps, and websites are interactive products to a certain extent, video games are, beyond any doubt, the epitome of interactivity. For this reason, the games themselves must provide high flexibility to allow the player to customise them according to their personal choices, which means that "[v]ideo game programming must take into account syntactical and morphological rules in order to phrase exchanges with players correctly" [4, p. 67]. Additionally, with a view to creating the suspension of disbelief and providing an immersive experience, the player is also given the possibility of selecting the main character's name, gender, race, etc. This is done by the use of variables, which are references to storage locations that contain words and are automatically replaced by the programme in order to complete a sentence. An example would be the phrase "She/He/They entered the room" or simply "% entered the room", in these cases, the system will provide the option that matches the player's preferences. Evidently, this poses innumerable constraints to translators who have to devise natural solutions while sometimes resorting to controlled language that avoid "[e]rrors in the rendering of these sentences in other languages - because of gender and number agreement, mode of address, etc." [4, p. 67] that would disrupt the abovementioned immersion. Furthermore, in the case of video games, the "fun" dimension of the field adds an extra layer of complexity to the translation process as the localisation and inclusion of cultural references becomes crucial to the success of the game itself. Historically, the first example of video game localisation that prioritised this precise aspect and tilted the focus of the translation towards the targeted audience was the adaptation of the title as well as the names and nicknames of the ghosts of the 1980 world-famous game *Pac-Man* (Image 1).

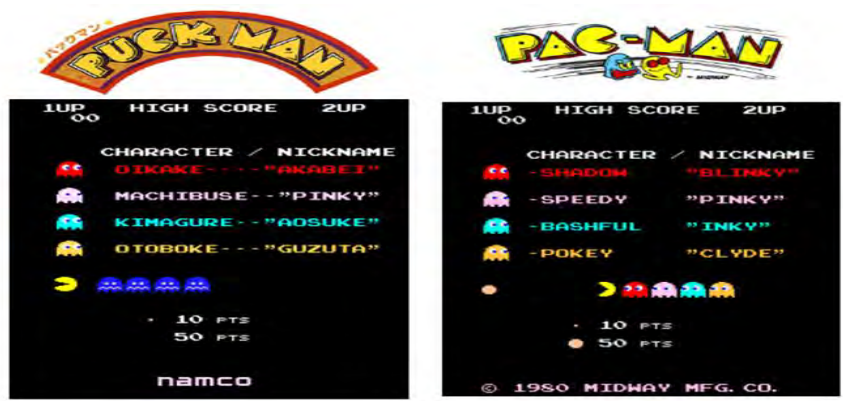

*Image 1: Japanese and US version of Pac-man [4, p. 226]*

In this particular case, the game was directly developed using English—or English spelling in the case of the ghosts—due to the fact that at the time Japanese characters could only be displayed as images. However, the company in charge of releasing the game in the US "considered necessary [to change the title] because the word "Puck" would likely tempt vandals in the US to slightly alter the first letter" [5, p. 49]. Moreover, the names of the characters were adapted into more edgy options in order to appeal to the US market, which demonstrated "the importance of pithy and punchy-sounding renditions, even to the point of choosing entirely new names in the target text (TT). This reflected […] the product's ultimate goal of providing some amusement to the end player" [5, p. 49]. Subsequently, other companies such as Nintendo of America started adapting the names of the characters as in some cases, due to cultural differences, their Japanese version did not use proper nouns. These practices introduced the concept of "transcreation", which has been described as "translation by invention" [5, p. 54] or "a translation that completely tilts the balance towards the target audience but claims to be the same product, despite those differences" [6, p. 34]. Therefore, transcreation became intrinsic to the field as a means to adding to the humour of the game or adapting puns, allowing professionals to greatly detach themselves from the source text and providing them much more freedom. The following table displays some examples of transcreation extracted from the games *Donkey Kong* (1981) and *Pokémon* (1996) regarding the names of the characters (Table 1).

| **GAME** | **JAPANESE** | **ENGLISH** | **FRENCH** |
|---|---|---|---|
| Donkey Kong | Jump-Man | Mario | Mario |
| Donkey Kong | The Lady | Polly | Polly |
| Pokémon | Hitokage | Charmander | Salamèche |
| Pokémon | Zenigame | Squirtle | Carapuce |
| Pokémon | Fushigidane | Bulbasaur | Bulbizarre |

*Table 1: Transcreation of character's names*

**Current business practices**

However, as Mangiron Hevia and O'Hagan point out, due to the lack of standardised practices in the field and the high number of *amateur* translators, "such liberties [transcreation] were sometimes also taken out of desperation rather than as a creative addition" [5, p. 54]. As the quality of the translations improved and the industry implemented internationalisation practices, the companies were able to release multilingual products without having to wait for the end of the development phase, thus shifting from a post-gold release model to a simultaneous shipment model. Sim-ship, therefore, entails translating the video game while it is still being developed, a practice that dramatically decreases access to reference material and the visual environment. Previous research in the field in the form of surveys shows that almost 70% of video game localisers have to deal with simultaneous shipment and that "80% of them stated receiving constant modifications during the duration of the project" [7, p. 40]. In a MasterClass imparted as part of a LocJam event organised by the University of Burgundy, Pierre Techoueyres explained in his capacity of professional video game translator and former QA project manager that "[constant changes] is something that happens quite often. They can be the result of a decision taken by the studio where they decide to rewrite a part of the story because they realised that a quest wasn't working and that affects translation or they can also derive from lack of information" [8]. Furthermore, whereas 85.16% of the 620 participants normally received Excel files with the text, only 14.84% indicated having access to the game itself, which leaves most of them working blind. In order to compensate for this lack of visual access, localisers are usually provided with reference material which includes different types of videos such as walkthroughs or promotional material (14.65%), screenshots of the game (12.82%), various types of images (12.45%), glossaries (10.26%), etc. The following table (Table 2) portrays some of the unedited comments left by the participants in regards to the type of reference material they usually received.

| 7/23/2020 5:35 PM | Game Design Documents, character bios, screenshots, videos... |
|---|---|

| 7/23/2020 10:03 AM | source files, videos, pictures, characters bios, SOPs and such |
|---|---|
| 7/30/2020 9:25 PM | It depends. Sometimes there is even access to the game. Some reference materials are always available. But as a rule there is not that much of it as I'd like to have. A usual practice are questions and answers (e.g. in a Google sheet). |
| 7/30/2020 7:12 PM | Public info like Store page, trailers and let's plays. Sometimes refs are provided by the devs: lore, character, mechanics descriptions, item pictures, etc. |
| 7/30/2020 4:11 PM | Game development files, sometimes a few pictures and some context about the game, but not much and most of the time only the official website |
| 7/30/2020 4:02 PM | Well prepared localization projects gives some images/videos as reference, as well as a word file explaining the backstory of the game. The best is when they provide an excel file with all the text in different tabs, in the right order. With the name and gender of the speaker as well as the name and gender of the person they are talking too. That is very rare though. |

*Table 2:* comments about reference material

Whereas the two first examples enumerate some of the files or assets that localisers usually receive in general, the third comment also points out a common practice of the field, the use of Q&A procedures such as Google sheets, emails, etc. Regrettably, as Pierre Techoueyres [8] explained miscommunication issues are quite common and can result in last-minute changes such as the case of a colleague who had to rewrite ten thousand words because he had either used "vous" instead of "tu" or vice versa. The following two comments express the fact that the professionals of the field need as much information and context as they can get and the fact that current practices are not enough. Finally, in addition to providing the type of material that a properly prepared localisation kit tends to include and insights into the specificities of their profession, the last comment highlights another obstacle in video game localisation: the lack of text linearity and the difficulty it generates when determining the gender of the speaker or the addressee. Although according to the results of the survey, the absence of internal cohesion (Figure 1) is not as common as the lack of visual environment, the combination of these two phenomena leads to what Bernal-Merino calls "a double-blind process (no audiovisual context, no text linearity)" [4, p. 117] and is amplified by the use of gender-related variables. All these factors contribute to the fact that mistranslation is the second most common type of linguistic bug found by the LQA (linguistic quality assurance) during the testing phase of the video game, which is carried out once all the localised strings have been integrated into the game.

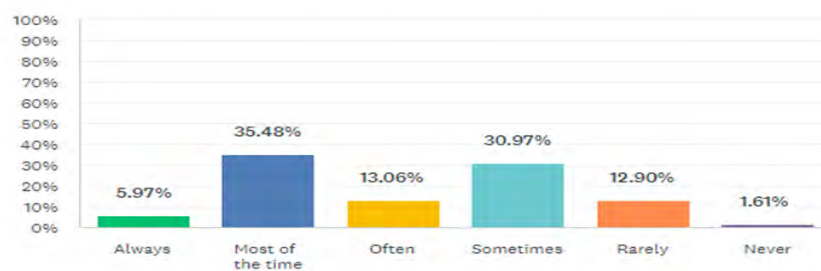

*Figure 1:* Text linearity in video games

# Working around gender issues in video games

      From this point forward we will focus solely on analysing the various techniques used by professional video game localisers as well as fan translators for accurately translating gender in the field of video game localisation, a complicated task in many cases as seen in the previous section. In order to provide reference material as well as to disambiguate the gender of the character in question, some development studios provide translators with character bibles, which are documents created by the design team to catalogue the name, gender, description, and biography of each *dramatis persona* for the rest of the teams and that can also be used by translators. Nevertheless, if the development team does not include in the Excel file the identity of the character speaking in that precise string, it is still common to find passages where, due to the nature of the text (i.e. lack of visual access, context, and text linearity), the source text does not clearly allow the translator to assess the gender. In the case of the translation from English into French, we can observe the cases of the pronouns "you" and "they" (and "it" as well) which are particularly problematic due to the fact that they can be translated as "tu" or "vous" in the first case or "ils" and "elles" in the second case. Whereas the latter depends on gender, the former entails discerning whether the original uses singular or plural as well as the use of the "vous" as a politeness formula. Additionally, the use of "vous" in its polite form allows the translator to avoid specifying the gender, similarly to the use of the French neutral formulation "on". The following table (Table 3) showcases examples of "vous" extracted from parallel corpora compiled from video games for the purpose of analysing gender-related translation strategies.

| VIDEO GAME | ENGLISH | FRENCH |
|---|---|---|
| The Witcher 3: Wild Hunt | Read the letter you found. | Lisez la lettre que vous avez trouvée. |
| The Witcher 3: Wild Hunt | If you hear hooves beating at night, but no one rides down the road... | Si, une nuit, vous entendez une cavalcade alors qu'aucun cheval n'est visible sur la route… |
| The Witcher 3: Wild Hunt | If you can see a blood-red glow on the horizon, but the air is free of smoke... | Si vous distinguez une lueur rouge sang à l'horizon alors que l'air ambiant est exempt de fumée… |
| The Witcher 3: Wild Hunt | What is magic? I can show you... If you're not afraid, that is. | Qu'est-ce que la magie ? Je peux vous montrer… si vous n'avez pas peur. |
| Deltarune | YOU MUST CREATE A VESSEL | VOUS DEVEZ CRÉER UN RÉCEPTACLE |
| Deltarune | YOU HAVE CREATED A WONDERFUL FORM | VOUS AVEZ CRÉÉ UN ÊTRE MAGNIFIQUE. |
| Deltarune | (You drank from the water fountain.) | (Vous buvez de l'eau de la fontaine.) |

***Table 3:*** *Use of the polite "vous"*

The table shows the examples that have been extracted using the parallel concordance option of Sketch Engine. In order to retrieve them, we looked for all the segments where "you" had been translated as "vous". The first example of *The Witcher 3: Wild Hunt* (*TW3*) and all of those extracted from *Deltarune*, show the strategy of using the singular pronoun "you" to directly address the player itself and, as such, it is necessary to avoid gender references. Although the previously mentioned "fun" factor and casual atmosphere of video games result in favouring the use of "tu" instead of "vous", in these cases the choice was made to create an immersive experience for players from all genders. The second and third examples of *TW3* are part of passages that can be found in a series of books that can be read by the main character. As such, the author of these virtual books addresses directly the audience (which is, in fact, the player itself although disguised as an unknown reader) and, in French, the translator decided to use impersonal sentences as a way of including all possible options. The final example from *TW3* simply shows an occasion where the source document does not provide enough context to identify the gender of the addressee and, thus, it has been translated as "vous" to avoid potential gender-related issues. As previously mentioned, besides the use of the polite "vous" in its singular form, the French language also proposed the alternative of re-writing the sentence in order to use the gender-neutral pronoun "on". The following table (Table 4) puts together some examples of the use of this technique extracted from our corpus.

| **VIDEO GAME** | **ENGLISH** | **FRENCH** |
|---|---|---|
| The Witcher 3: Wild Hunt | You wanna sail to Skellige, eh? | Alors comme ça, on veut voguer jusqu'à Skellige ? |
| The Witcher 3: Wild Hunt | Unexpected child? Aye, have one of those each year… Unless you mean something else? | L'enfant surprise ? J'en pond un par an, je connais… A moins qu'on parle pas de la même chose ? |
| The Witcher 3: Wild Hunt | They milk you dry… | On se fait saigner à blanc |
| The Witcher 3: Wild Hunt | While in the passageway's underneath Novigrad (I've been there as well and don't recommend tourist destinations - if the stench doesn't get you, the monsters will.) | Dans les souterrains qui courent sous Novigrad (pour m'y être rendu, je le déconseille aux touristes : on peut survivre à la puanteur, plus difficilement aux monstres qui y vivent.) |
| The Witcher 3: Wild Hunt | They'll wash your clothing for you there. | On pourra y laver vos vêtements. |
| The Witcher 3: Wild Hunt | Such is life – you can't win them all, sadly. | Ainsi va la vie. On ne peut pas toujours gagner, hélas. |
| Alt-Frequencies | But is it? | En est-on sûr ? |
| Alt-Frequencies | They feel that information is being withheld from them. | Ils ont le sentiment qu'on ne leur dit pas tout. |

**Table 4:** *Use of "on" for rephrasing*

If we analyse the examples in the previous table we can observe that the first and second rows show the use of "on" to avoid having to decide between using "tu" and "vous" due to lack of context. This decision has been taken because each non-playable character (NPC) of the games addresses Geralt differently and the translator has to choose every single time. The rest of the examples show the use of reformulation in order to include both genders as the text does not provide enough information. In particular, examples number four and number six, besides to avoid gender-related issues, have been used to deal with narrative discourse. Whereas the former, once again, can be found in the form of a passage of a book, the latter is a sentence said by Jaskier, a non-playable character who also narrates the game. The segment enounces a fictive universal truth which, in French, is generally done via the use of the pronoun "on". Finally, the last two examples show the classic technique of rephrasing to use "on" in order to avoid gender-related issues in a situation that requires extra information.

## The emergence of non-binary characters

Nowadays, we can observe the appearance of non-binary characters in the media and a societal trend towards inclusivity and breaking down the binary conception of gender. This widespread movement has, evidently, also reached the video game industry and developers have started to introduce non-binary characters in their games. Nevertheless, "[w]hile it might be the case that there is good non-binary gender representation in some other games, especially games developed by queer indie game developers [...], this does not always appear in AAA games" [9, p. 227-228]. Therefore, as a consequence, translators will have to base their decision on probabilities if they are not provided with the proper gender in a comment inserted in the Excel file. As Techoueyres explained in his interview:

"Personally, I worked on a project a couple of years ago where one of the main characters was a lady that wrote letters to the person she loved, and during six months we were wondering whether it was a man or a woman. So we interpreted it as a man since 99% of the time it is the case. However, two months before the release of the game, during the testing phase, the studio wrote back and told us that it was actually a non-binary character." [8].

In the light of these changes and in order to analyse the strategies used by video game localisers when translating from English into French, we decided to create a specific corpus with games that included female leading characters as well as transgender, non-binary, and non-sexualised characters. Furthermore, we decided to use games that had been translated by humans from the beginning (favouring official translations), as it has become common in the industry to resort to machine translation (MT) and MT postediting in order to reduce the costs. This practice, which is not exclusive of the field, is reinforced by the introduction of neural machine translation (NMT) systems directly into the game development engine, the explosion of the video game market, and the subsequent increase in competitiveness. However, previous research in the field of video game localisation has proven that these tools can cause up to 3,03% of mistranslations directly related to gender bias [10] in a field where, as previously explained, suffers from chronic issues when it comes to identifying gender.

The parallel corpus has, for now, over 500.000 words in English and more than 600.000 in French and includes games such as *Alt-Frequencies*, *A normal Lost Phone*, *Industries of Titan*, *Crypt of the NecroDancer*, *Deltarune*, or *The faces of the forest*. All these games have at least one non-binary character with the exception of *A normal Lost Phone*, which has a transgender main character but, due to the nature of the game, parts of it use neutralisation techniques in order to avoid revealing this fact to the player. Moreover, in the process of creating this corpus, we observed that many of the Excel files in the case of indie game developers included comments and instructions related to the gender of each character or at least highlighted the presence of each non-binary character (although some of them were not consistent or very clear). The following table (Table 5) displays some examples of neutralisation techniques extracted from said games.

| VIDEO GAME | ENGLISH | FRENCH | NEUTRALISATION TECHNIQUE |
|---|---|---|---|
| Deltarune | Kris! Show up earlier next time. | Kris ! Tu devrais te lever plus tôt. | Changing the tense to create an impersonal structure. |
| Deltarune | Kris! don't act shocked. You know it's true | Kris, ne prends pas cet air surpris. Tu sais que c'est vrai. | Replacing the adjective with a noun. |
| Deltarune | Let's go freak | Dépêche-toi, minable | Omitting the article. |
| Alt-Frequencies (Charlie) | I have to be honest; I am intrigued. My science-heart is beating faster, so to say. | Je dois l'avouer, je suis perplexe. Ma curiosité scientifique a été piquée, si je puis dire. | Using "perplexe" instead of "intrigué(e)" to avoid gender agreement issues. |
| Alt-Frequencies (Charlie) | Just to correct you, I am not an expert on the time loop. | Précision, je ne pense pas pouvoir m'attribuer une quelconque expertise en boucle temporelle. | Rephrasing to avoid mentioning the character's gender. |
| Alt-Frequencies (Charlie) | I'm not giving out advice on how to vote if that's what you're looking for. | Si vous me demandez de donner une consigne de vote, je m'y refuse. Mon domaine c'est la science, pas la politique | Restructuring the sentence. |
| Alt-Frequencies (Charlie) | Let me put it this way. Think of our report as a medication information leaflet. | Il faut le voir comme une brochure d'information médicale. Une brochure provisoire, parce que... | Deleting a full sentence. |
| Alt-Frequencies (Michelle) | I'm already fed up with the subject anyway. | Ça m'énerve de toute façon | Rephrasing and condensing the information. |
| Alt-Frequencies (Fred) | I feel like a fool. | Je me sens bête. | Using a term that suits both females and males. |

*Table 5:* *Neutralisation techniques*

The examples in the table above are those of techniques that can be categorised into three main strategies. The first method—which is the one employed in the case of rows number one, two, five and eight—consists of rephrasing the whole sentence in order to avoid any reference to gender by either completely changing the whole sentence (see examples five and eight) or part of it (examples one and two). To this end, the modification can focus on a lexical level, such as in the case of the second example that shows a shift of the gender agreement from "Kris" to the French word "air", or at a tense level, where the syntax and verb tense has been changed to create an impersonal structure. The second strategy—illustrated by examples number three, four, and nine—aimed to employ nouns or adjectives that suit both males and females or have the same spelling to neutralise the sentence. Finally, the third strategy entails deleting or restructuring parts of the sentence to remove any gender references as demonstrated in example number six and example number seven.

The strategies analysed in the previous paragraph are frequent when it comes to neutralising gender, but not specific to inclusive language and have been applied for years in the translation field [11], and especially in localisation. As society evolves to become more inclusive, languages also adapt to reflect these changes, albeit at a slower pace. Originally, French inclusive language - defined as "a set of graphic and syntactic measures to ensure equal representations of women and men" [12] - was developed to reduce inequality between men and women. At first, there were three conventional rules to follow to reduce biases: the use of both the feminine and masculine forms; the gender agreement of status, professions, and titles; and to avoid antonomasia such as the use of "Femme" and "Homme". However, those rules only targeted the topic of equality and fairness between men and women and, as a consequence, it is more appropriate to label these techniques as gender-fair language. As a matter of fact, these first attempts at a French Inclusive Language were not well received and were subject to complaints due to the exclusion of the LGBTQA+ community.

As a result, other methods were added to the previous three for inclusivity reasons [13] such as the creation of new grammar rules, inclusive pronouns, possessives and adjectives, new spellings, and neutral-gender agreements [14]. These are still experimental and not yet anchored in the language, and have been branded as "awkward", "unusual", or difficult to read or understand. Therefore, the topic is surrounded by controversy and there are many debates associated with socio-political and linguistic approaches around the use of French inclusive language. Furthermore, in the field of video games, it has been argued that the use of these alternatives could negatively affect the attention of the player, thus breaking the suspension of disbelief and hindering the product's immersive nature. Nevertheless, it is a trend that seems to be in vogue and, with time, French-speaking players might become accustomed to these strategies or even come to expect them depending on the game.

In our corpus, we can find examples of these newly created inclusive rules, which seem to be gaining popularity among indie game developers from the LGBTQA+ community. The first example that we can observe is the technique of paraphrasing and re-writing the sentence in order to avoid indicating the gender or to use terms that include all genders. This is the case of the example extracted from *Industries of Titan* and showcased in the following table (Table 6) where "enfant" has been used to avoid the nouns "fille" or "fils". Nevertheless, the use of recent gender-inclusive pronouns and possessives such as "iel", "iels", "ses", "lea" seems to have the highest number of occurrences within our data, which points to an emerging trend in terms of relevance and representativeness. Table 6 shows various examples of most of them in context. Finally, the last example displays the use of gender-fair language in the form of including both the feminine and the masculine forms and separating the extra "e" used for the agreement by a dot.

| VIDEO GAME | ENGLISH | FRENCH | TECHNIQUE |
|---|---|---|---|
| LongStory | Seriously? There's a person inside that lice | Sérieux ? Il y'a une personne dans ce truc | Using the inclusive language with the French |
| | farm. And they've been wearing that costume since the first day of school last year. Last YEAR. Do you understand how gross that is? They probably shower in it. | dégueu. Et iel porte ce costume depuis la rentrée de l'année dernière. L'année DERNIERE. Comprends-tu à quel point c'est dégoûtant ? Iel prend sûrement sa douche avec. | gender-neutral pronoun singular "iel". |
| LongStory | "I'm kidding, they just had to go." | "Je rigole, iel a juste dû partir." | Using the inclusive language with the French gender-neutral pronoun singular "iel". |
| Industries of Titan | Brought up by the founders of the small but successful Coroba Solutions Bahar witnessed how Titan's corporations treated its employees. When they spoke out about it their parents tried to have them Converted. That was enough to push them towards the rebel cause. | Enfant des fondateurs de la petite mais prospère Solutions Coroba Bahar a été témoin dès son plus jeune âge du traitement réservé aux employés par les entreprises de Titan. Quand iel a abordé le sujet ses parents ont essayé de lea faire convertir. Ce fut suffisant pour l'inciter à rejoindre la cause rebelle. | Using the neutral term "enfant" for the translation of "brought up by". Then, following the developers' instructions, "they"; "their" and "them" have been translated by iel", "ses", "lea" respectively. |
| Alt-Frequencies (Fred to Charlie) | Professor Thomas, you have been researching the effects of a possible time loop on public health, am I correct? | Professeur·e Thomas, vous avez effectué des recherches sur l'impact possible d'une boucle temporelle sur la santé, n'est-ce pas ? | This segment used another strategy of inclusive writing in French that consists of including both the feminine and masculine forms separating them with a dot. |

*Table 6:* Gender-inclusive techniques

## Conclusion

This article has presented the preliminary results issued from a parallel corpus created with a two-fold purpose: to catalogue and extract neutralisation techniques used by translators and, ultimately, to use the data to train a neural machine translation system. The project, financed by ISITE BFC and the programme « Investissement d'Avenir », aims at creating the first NMT tool specialised in neutralisation techniques applied to the field of video game localisation, hence the name All-inGMT. Moreover, the field of video games was chosen due to the already present ambiguity of the industry and the characteristics presented during the first part of the article, which makes video game localisation ideal for identifying neutralisation techniques.

**References:**

1. UREN, E., HOWARD, R., & PERINOTTI, T. *Software Internationalization and Localization: An Introduction.* 1993. ISBN 10: 0442014988
2. DUNNE, K.J. Localization and the (R)evolution of Translation. En: S. BERMANN, C. PORTER ed. *A Companion to Translation Studies.* John Wiley & Sons, Ltd. 2014, p. 147-162.
3. PYM, A. *The Moving Text: Localization, translation, and distribution.* John Benjamins Publishing Company, 2004. ISBN 9789027295828
4. BERNAL- MERINO M.A. *The Localisation of Video Games.* Translation Studies Unit Imperial College, London, 2013, p. 67.
5. MANGIRON HEVIA, C. & O'HAGAN, M. *Game Localization: Translating for the global digital entertainment industry.* John Benjamins, 2013. ISBN 9789027271860
6. BERNAL- MERINO M.A. On the Translation of Video Games. In: *JoSTrans, The Journal of Specialised Translation.* 2006. No. 6. p. 22-26. ISNN 9781315752334
7. RIVAS GINEL, M.I. Ergonomics of Tools Usage for Video Game Localisation: A User Survey. In: >*Critic*. 2021. No.2. p. 27 – 57. ISBN-13: 979-8453471812
8. TECHOUEYRES, P. *LocJam Masterclass*. 2020. min. 27:11 – 28:47 (consulted the 13.03.2022) available at https://www.youtube.com/watch?v=YAbl257stHo&ab_channel=LocalisationDijon
9. HERITAGE, F. *Language, Gender and Videogames*. Switzerland: Palgrave Macmillan, 2021. ISBN: 978-3-030-74398-7.
10. RIVAS GINEL, M.I. & THEROINE, S. Machine Translation and Gender biases in video game localisation: a corpus-based analysis In: *Colloque interdisciplinaire « Vers une robotique du traduire ? »*, 30 sept. – 1er oct. 2021. Strasbourg. ⟨hal-03540605⟩
11. GOTTLIEB, H. A New University Discipline. In: C. DOLLERUP, & A. LODDEGAARD, ed. *Teaching Translation and Interpreting: Training, talent and experience*. John Benjamins Publishing Company. 1992. DOI: 10.1075/z.56.26got
12. HADDAD, R. *Manuel d'écriture inclusive.* 2019. (consulted the 14.03.2022) available at https://www.motscles.net/ecriture-inclusive
13. CREMIER, L. Rédaction inclusive, féminisation et approches créatives du genre grammatical en français: comment traduire le genre. In : *4ème conférence du Comité pancanadien de terminologie. Équité, diversité et inclusion : l'importance de la terminologie.*, 27 janvier 2022. Online. UQAM, IREF, 2022.
14. ALPHERATZ. *Grammaire du français inclusif: littérature, philologie, linguistique*. Ed: Vent solars, 2018. ISBN: 9782955211861